\documentclass[prl,twocolumn,showpacs,amsmath,amssymb]{revtex4}
\usepackage{amssymb}
\usepackage[dvips]{graphicx}

\begin{document}

\title{Current-voltage characteristic of narrow superconducting wires: bifurcation phenomena.}
\author{ V. V. Baranov$^{1}$, A. G. Balanov$^{2}$, and V. V. Kabanov$^{1}$}

\affiliation{$^{1}$Jozef Stefan Institute, Jamova 39, 1001 Ljubljana,
Slovenia; $^{2}$Department of Physics, Loughborough University, Loughborough, UK}

\begin{abstract}
The current-voltage characteristics of long and narrow superconducting channels are
investigated using the time-dependent Ginzburg-Landau equations for complex order parameter. We found out that
the steps in the current voltage characteristic can be associated with bifurcations of either steady state
or oscillatory solution. We revealed typical instabilities which induced the singularities in current-voltage
characteristics, and analytically estimated period of oscillations and average voltage in the vicinity of the
critical currents. Our results show that these bifurcations can substantially complicate dynamics of the order
parameter and eventually lead to appearance of such phenomena as multistability and chaos.
The discussed bifurcation phenomena sheds a light on  some recent experimental findings.

\end{abstract}

\maketitle

The problem of the appearance of the resistive state in narrow
superconducting channels when the current $j$ exceeds the value of the
critical current $j_{c}$ was the subject of intensive debates in
the past\cite{IvlevKopnin}. The voltage between two parts of the
superconducting wire appears when the superconducting order parameter (OP)
is strongly suppressed near the phase slip center (PSC). The phase of the OP
becomes very steep function across the PSC and finally
demonstrates $2\pi$ jump exactly when the OP reaches 0\cite{IvlevKopnin,ludac}.
This process appears periodically in time leading to
the oscillation of the OP and nonlinear current-voltage characteristic (CVC)
with well pronounced steps of voltage when $j$ reaches certain
critical values\cite{Skocpol,Dmitrenko}.

The PSC has been extensively investigated
theoretically and experimentally in the past(see Ref.\cite{IvlevKopnin}
and references therein). It is well established that if
$j>j_{2}\sim j_{c}l_{E}^{2}/\xi^{2}$ the normal state is absolutely stable,
here $l_{E}$ is the penetration depth of the electric field into
the superconductor and $\xi$ is the superconducting coherence length
Therefore if $l_{E}>>\xi$ there is a wide interval of currents $j_{c}<j<j_{2}$ where
the superconducting state coexists with the normal state leading to the
oscillating type of behaviour\cite{KramerWatts,KramerBaratoff}.

The steps in the CVC are usually interpreted as points where new PSC penetrates
to the wire\cite{IvlevKopnin,Skocpol,Dmitrenko}. It was also demonstrated that
the time average of the voltage $V$ and the frequency of
oscillations $\omega$ satisfy the relation similar to the Josephson
relation $V\propto \omega$ \cite{IvlevKopnin}. The theoretical interpretation of
the CVC is usually based on the statical model of the resistive
state\cite{Skocpol,Galaiko} or a very similar dynamical theory of the resistive
state\cite{IvlevKopnin,IvlevKopninMaslova}. All these models are based on
the approximate time averaging of the time dependent Ginzburg-Landau equations(TDGLEs) over
a period. Therefore the complex behaviour of the order parameter during one
cycle is lost in these theories. An important step forward in the analysis of the TDGLEs
has been made recently \cite{mic,vod}. It has been found that the dynamic of the order parameter
in the PSC is governed by the ratio between the relaxation times of the TDGLEs\cite{mic,vod,ludac}.
It has been established that the
CVC exhibits an S shape in the regime of the fixed voltage. Besides, it has been shown that in the
limit of strong pair breaking effect due to interaction with phonons the CVC demonstrates strong hysteresis
in the regime of constant current\cite{mic}. This behaviour appears because two different solutions may
coexist at the same value of the current. All these results (see also \cite{brd1,brd2})
demonstrate that different periodic and quasiperiodic in time solutions emerge with the changing of the
voltage and current\cite{kim}.
However, some of the properties of these solutions are still not understood.
The types of bifurcations when one solution is replaced by the other solution are not identified.
The theory still lacks of an accurate analytical description of the CVC and well defined
and experimentally verifiable predictions. In this paper we present a detailed study of the CVC
in the resistive state of superconductors with different lengths. We reveal the types of
bifurcations corresponding to the discontinuities on the CVC and derive an analytical expression for
CVC when $j \to j_{c}$. We predict the appearance of the
frequency in the spectrum of the electromagnetic radiation generated by the current after the period doubling
bifurcation. This frequency equals exactly to the half of the one before the bifurcation point.
Besides, we identify the positions of the PSCs in the wire and prove that our predictions
on the basis of TDGLEs are in excellent  agreement with recent experimental results of
Sivakov \emph{et al.} \cite{Sivakov}.

Recent experimental studies have detected the spatial positions of PSCs in the different
parts of the CVC\cite{Dmitrenko,Sivakov}. With the help of low
temperature laser scanning microscopy technique it was demonstrated that
each quasilinear part of the CVC has different number of the PSCs.
When the current becomes larger then $j_{c}$ the first PSC appears in the
middle of the wire. Further increase of the current leads to the second step
in voltage. Two PSCs appear symmetrically with respect to the center of the wire.
Further increase of the current leads to the the appearance of the third step in
voltage and the third PSC is created in the center of the wire
(see Fig.2 of Ref.\cite{Sivakov})

Another important effect was observed recently on the CVC of submicron BSCCO bridges\cite{Zybtsev}.
When the temperature of the bridge is reduced bellow $T_{c}$, characteristic steps on the CVC appear
with clearly observed hysteresis. The measurement of the voltage as a function
of time at high temperatures and at the fixed current has demonstrated sharp switching between two metastable states.
The voltage randomly jumps between two values (random telegraph noise)\cite{Zybtsev}.
The average frequency of the switching is very strong function  of temperature: decreasing of temperature by 1 K
leads to the decrease of the frequency of switching by several orders of magnitude.  Therefore the
activation energy that corresponds to the switching is more then $10^{4}$K\cite{Zybtsev}, indicating
the macroscopic nature of the barrier between metastable states.

In our work, which involves
numerical and analytical analysis of TDGLEs, we show that
each step of the CVC corresponds to the bifurcation point where the solution looses stability
and a new topologically different solution emerges. At $j=j_{c}$
the
steady state solution looses stability as a result of the saddle-node homoclinic
bifurcation \cite{Kuznetsov}, which leads to the birth of a limit cycle of infinite period. With the further
increase of $j$ this limit
cycle looses its stability leading to the period-doubling bifurcation.
This behaviour is universal and does not depend on the length $L$. The
next bifurcation point is not universal and depends on $L$. With the
further increase of $L$ we observe the effect of multistability. Two or more different limit
cycles coexist at the same value of the current.

The first TDGLE in dimensionless units has the form:
\begin{equation}
u(\frac{\partial\psi}{\partial t}+i\phi \psi) =
\frac{\partial^{2}\psi}{\partial x^{2}}+\psi-
\psi|\psi|^{2}.\label{tdgl1}
\end{equation}
The second equation defines the stationary current through 1D wire:
\begin{equation}
j=-\frac{\partial\phi}{\partial x}+\frac{1}{2i}
(\psi^{*}\frac{\partial\psi}{\partial x}-
\psi\frac{\partial\psi^{*}}{\partial x})\label{curr1}.
\end{equation}
Here $\psi$ is the dimensionless complex OP, the distance is measured
in units of the coherence length $\xi$ and time
is measured in units of the phase relaxation time
$\tau_{\theta}=\frac{4\pi\lambda^{2}\sigma_{n} }
{c^{2}}$, $\lambda$
is the penetration depth, $\sigma_{n}$ is the normal state
conductivity, and $c$ is the speed of light. The parameter
$u=\frac{\tau_{\psi}}{\tau_{\theta}}$ is a material dependent
parameter, where $\tau_{\psi}$ is the relaxation time of the
amplitude of the OP. The electrostatic potential $\phi$ is measured in the units of
$\phi_{0}/2\pi c\tau_{\theta}$, where $\phi_{0}=\pi \hbar c/e$ is the flux quantum, $e$ is the electronic charge and $\hbar$
is the Planck constant, the dimensionless current density $j$ is defined in
the units of $\phi_{0}c/8\pi^{2}\lambda^{2}\xi$. Nonequilibrium superconductors are
characterized by an additional length scale $l_{E}$ which describes the penetration
of electric field. From Eqs.(\ref{tdgl1},\ref{curr1}) it follows
that $l_{E}=\xi/u^{1/2}$. The time-dependent Ginzburg-Landau equations for complex order
parameter were derived for $T<T_{c}$ in late 60th \cite{Eliashberg,AbrahamsTsuneto}. The
simplified version of the TDGLEs used in this work corresponds to the so-called gapless case with large
concentration of magnetic impurities $\tau_{s}T_{c} <<1$, where $\tau_{s}$ is the relaxation
time on magnetic impurities \cite{KopninKniga}. In that case parameter u is 12. More realistic
generalized TDGLEs are derived in dirty limit $\tau_{im}T_{c} <<1$, where $\tau_{im}$ is impurity
scattering time. These equations take into account inelastic scattering by phonons.
In that case parameter u is 5.79. The penetration depth of the electric field $l_{E}$ is substantially
enhanced if the inelastic scattering time is large enough. The enhancement of $l_{E}$ may be approximated
in the case of gapless superconductivity if we assume that phenomenological parameter $u$ is small
$u<1$\cite{IvlevKopninMaslova}. This approximation is widely accepted in the literature and leads
to qualitatively and often quantitatively correct results above and below $T_{c}$.
Here we consider the superconducting wire of a
length $L$ with the following boundary conditions:$\rho(-L/2)=\rho(L/2)=1$ and
$d\phi(-L/2)/dx=d\phi(L/2)/dx=0$, where we define phase $\theta$ and modulus $\rho$ of
the order parameter according to the relation $\psi=\rho\exp{(i\theta)}$. The absence of
the electric field at the end of the wire
determines the gradient of phase $d\theta(-L/2)/dx=d\theta(L/2)/dx=j$.

We start with the analysis of the
steady state solutions of the TDGLEs (\ref{tdgl1},\ref{curr1})
 which are defined as $\psi=(1-k^{2})^{1/2}\exp{(ikx)}$. The equation for $k$ has the form
$j=(1-k^{2})k$, and for $j<j_{c}=2/3\sqrt{3}$ it
yields two roots $k_{1} < k_{2}$. The first
solution corresponding to $k_{1}$ is stable and the second ($k_{2}$) is unstable.

 When
the current reaches its critical value $j=j_{c}$ these
steady state solutions collide with each other ($k_{1}=k_{2}$) and
disappear, thus suggesting a saddle-node bifurcation. Indeed, the the stability analysis of the
steady state solution performed
in Ref.\cite{ludac} shows that only one Lyapunov exponent crosses $0$ when $j=j_{c}$.
Additionally, our numerical calculation revealed that at the moment of this bifurcation,
a periodic solution (limit cycle) with very large period arises in the phase space of the
system in the very close vicinity of the disappeared steady states. As $j$ grows, the period of
the limit cycle quickly decreases. All those facts evidence that at $j=j_{c}$ the
systems undergoes the saddle-node homoclinic bifurcation \cite{Kuznetsov}.



Note that the value of $j_{c}=2/3\sqrt{3}$ is obtained for the wire of the infinite length.
The boundary conditions for Eqs.(\ref{tdgl1},\ref{curr1}) defined above lead to the slightly different
value of the critical current. Therefore the value of $j_{c}(L)$ should be determined from the
solution of stationary equations (\ref{tdgl1},\ref{curr1}) with the same boundary conditions.

According to the theory of dynamical systems,
the behaviour of a system in the vicinity of the saddle-node homoclinic bifurcation can be described by
the normal form\cite{Kuznetsov}
\begin{equation}
\dot{y}=\beta+a(0)y^{2},
\label{normal}
\end{equation}
where $y$ is a representative phase variable, $\beta=-2^{3/2}u(j-j_{c})/3(u+2)j_{c}$, $a(0)=-2^{3/2}u(u+2)$. To derive the normal form we
have to follow the standard recipe Ref.\cite{Kuznetsov}. It is convenient to rewrite the TDGLs
(\ref{tdgl1},\ref{curr1}) in the limit of $L\to\infty$ in terms of $\rho$ and the gauge invariant scalar
$\Phi=\phi+\partial\theta/\partial t$ and vector $Q=\partial\theta/\partial x$ potentials.
\begin{eqnarray}
u\partial\rho/\partial t&=&\partial^{2}\rho/\partial x^{2}+\rho(1-\rho^{2}-Q^{2})
\nonumber \\
u\rho^{2}\Phi&=&\partial (\rho^{2}Q)/\partial x \nonumber\\
j&=&-\partial\Phi/\partial x +\partial Q/\partial t +\rho^{2} Q
\label{gieq1}
\end{eqnarray}
These equations have the
steady state
solution $\rho=\rho_{0}=\sqrt{2/3}$, $\Phi=0$,
$Q=1/\sqrt{3}$, $j=j_{c}$, which becomes unstable and has one Lyapunov exponent $\lambda_{0}$ which
crosses 0 at $j=j_{c}$. Expanding Eq. (\ref{gieq1}) up to the second order
in deviations from steady state
solution, rescaling the time $t/u \to t$ and projecting this equation on the direction
of the eigenvector corresponding to $\lambda_{0}$ we obtain Eq.(\ref{normal}). The normal form
(\ref{normal}) allows to predict the period $T$ of the periodic solution which
corresponds to the limit cycle. Following to Eq.(32,23) of Ref.\cite{Landau} we
integrate equation (\ref{normal}) over $y$  between $\pm y_{1}$ where $y_{1}$ is of the order of $1$.
$T=2\tan^{-1}{(y_{1}\sqrt{a(0)/\beta})}/\sqrt{a(0)\beta}$. Therefore, when $(j-j_{c})/j_{c}\ll1$
the period of oscillations is determined by the formula:
\begin{equation}
T=\pi/\sqrt{a(0)\beta}=\frac{\pi\sqrt{3}(u+2)}{2^{3/2}u}\bigl[(j-j_{c})/j_{c}\bigr]^{-1/2}.
\label{period}
\end{equation}

To verify theoretical prediction we have performed extensive numerical simulations of
Eqs. (\ref{tdgl1}) and (\ref{curr1}), using the fourth order Runge-Kutta method. The spatial
derivatives are evaluated using a finite difference scheme of the second and the fourth order.
The calculations were performed for the superconducting wire of three different length
$L=L_{0}$, $2L_{0}$ and $4L_{0}$ ($L_{0}/\xi=10.88$) and for $u=1/2$. The different length leads to a slightly
different value of the critical current caused by the boundary conditions. The calculated
critical currents are $j_{c}(L)/j_{c}=1.016$, $1.002$ and $1.00019$ respectively. Some
properties of the solution are universal and do not depend on the length $L$. On the other hand,
there are some features of the solution and the CVC which are not universal and depend on the length
$L$.

\begin{figure}
\includegraphics[width = 60mm, angle=-90]{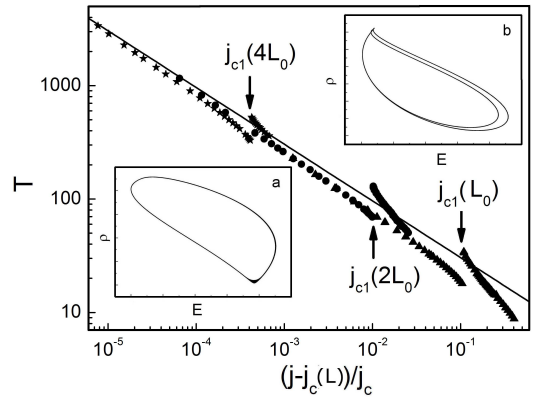}
\caption{The period of the solution as a function of current for three different lengths of the wire:
$L=L_{0}$(triangles), $2L_{0}$(circles), $4L_{0}$(stars). The solid line represents the results of Eq.(\ref{period}).
Arrows indicate the period-doubling bifurcation point $j_{c1}(L)$. Insets represent projection of the
limit cycle trajectory to the ($\rho(0),E(0)$) plane before (a) and after (b) bifurcation point $j_{c1}(L)$. Here
$E(0)$ is the electric field in the center of the wire.}
\end{figure}

The first universal property of the solution is the dependence of the period of the oscillation on
$(j-j_{c}(L))/j_{c}(L)$ in the range of currents $j_{c}(L)<j<j_{c1}(L)$, where $j_{c1}(L)$
is the first critical current where the limit cycle looses its stability. According to Eq.(\ref{period})
$T\approx 9.62\bigl[(j-j_{c}(L))/j_{c}(L)\bigr]^{-1/2}$ for $u=1/2$. The results of calculations
of the period for the oscillatory solution are plotted in Fig. 1 for all three lengths $L$
together with the analytical estimate,
given by Eq.(\ref{period}). There is very good agreement of numerical results with Eq.(\ref{period})
over more then four orders of magnitude in $(j-j_{c}(L))/j_{c}(L)$. Since Eq.(\ref{period}) describes
the asymptotic behaviour of the period $T$ in the limit $(j-j_{c}(L))/j_{c}(L)\to 0$ the agreement becomes better when
the current is closer to its critical value.

The divergence of the period implies that the CVC near $j_{c}(L)$ should
have nonlinear form $V \propto T^{-1} \propto \bigl[(j-j_{c}(L))/j_{c}(L)\bigr]^{1/2}$ (Fig.2). This type of
dependence is due to the quantization rule, derived in\cite{IvlevKopnin2}, that connects the averaged
electric field and the period of oscillations. Indeed, as it is clearly seen from Fig.2, the voltage $V$ shows
the universal dependence in accordance with Eq.(\ref{period}). Note that similar CVCs have been reported
earlier\cite{mic}. However, the authors did not recognize the type of bifurcation and did not derive the normal
form for this bifurcation. As a result they were unable to get analytical expression for CVC in the vicinity
of $j_{c}$. An accurate solution of dynamical equations
(\ref{tdgl1},\ref{curr1}) and asymptotic, predicted by Eq.(\ref{period}), do not confirm an
approximate result $V\propto-1/\ln{((j-j_{c})/j_{c})}$, derived in Ref.\cite{IvlevKopnin}. The square
root behaviour of the CVC, however, is in agreement with the earlier result of Aslamazov
and Larkin\cite{AslamazovLarkin}, who had considered the CVC of the short ($L\ll\xi$) superconducting bridge.

For relatively small currents $j_{c}(L)<j<j_{c1}(L)$
the phase slip center appears periodically in time in the center of the wire (Fig.2) in the agreement with
the experimental results\cite{Sivakov}. This oscillating
behaviour of the OP leads to the oscillating behaviour of the voltage and emission of an electromagnetic
radiation with the frequency $\omega_{1} \propto T^{-1}$.


\begin{figure}
\includegraphics[width = 60mm, angle=-90]{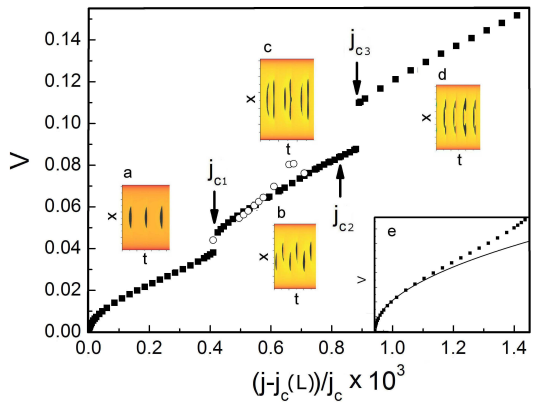}
\caption{ The CVC of the channel with $L=4L_{0}$. Open circles represent the more complex
limit cycles, which coexist with the main periodic solution (multistability). Insets (a-d) show the space-time arrangements
of the PSCs in different regions of the CVC. The inset (c) corresponds to open circles. Inset (e) shows the CVC
for $j<j_{c1}(L)$ in comparison with analytical formula (see the text).}
\end{figure}

Further increase of the current leads to
instability of the limit cycle. At $j=j_{c1}(L)$ the
periodic solution becomes unstable and
a new limit cycle emerges as the result of period-doubling bifurcation.
This is clearly seen from Fig.1 where
a sudden increase of the period is observed at $j=j_{c1}(L)$.
It appears that this instability is universal and does not depend
on the length $L$.

The period doubling bifurcation could be easily
recognized if we plot the cycle trajectories below and above $j_{c1}(L)$. It is important to plot
trajectories only for the gauge invariant quantities to avoid unphysical effects. Therefore, in the
insets to Fig.1 we present the periodic solution in coordinates $\rho(x=0)$ and electric field
$E(x=0)=-d\phi(x=0)/dx$. As clearly seen from this figure,
the period-doubling bifurcation manifests itself in the transition from a single-loop (period-1)
limit cycle to a double-loop (period-2) limit cycle \cite{Anischenko}.


This bifurcation also leads to the
singular behaviour of the CVC. This singularity is
more prominent
for longer wires $L=4L_{0}$
where the jump of voltage is well pronounced, while for $L=L_{0}$ it manifests itself
only by the change of the slope (Fig. 3 a). The arrangement of the PSCs in space and time is plotted in Fig.2.
As it follows from Fig.2, at $j=j_{c1}(L)$ the position of one PSC
is shifted up with respect to the center and the position of the next PSC is shifted down.
Therefore, the period of a new limit cycle is twice of the period of the limit cycle before the
bifurcation point. Each period includes now two PSCs, one is slightly above and another is slightly below the
center of the wire, which is well agreed  with experimental observations  \cite{Sivakov}.
As a result of period-doubling, a new frequency $\omega_{2}=\omega_{1}/2$
appears in the spectrum of the electromagnetic radiation generated by the current.
It should be pointed out that the period doubling bifurcation in the superconducting channel with fixed voltage
was reported earlier\cite{kim}. Nevertheless there is substantial geometric difference between this two cases. In
Ref.\cite{kim} the PSCs were always detected in the center of the wire, in contrast with our results.

Note that the critical current for period-doubling bifurcation $j_{c1}(L)$ is not universal and strongly depends on
the length of the channel $L$. According to Fig.3 $(j_{c1}(L)-j_{c})/j_{c}=0.125,0.012,0.0006$
for $L=L_{0},2L_{0}$ and $4L_{0}$ respectively. Indeed it demonstrates the exponential dependence on the
length $L$, as it was proposed in Ref.\cite{Galaiko1}.

The next bifurcation which is common for all studied cases is the destruction of
the periodic cycle. For all studied cases there exists the second critical current
$j_{c2}(L)/j_{c}=1.44,1.027$ and $1.001$ for $L=L_{0},2L_{0}$ and $4L_{0}$ respectively
where the periodic solution looses its stability. This transition does not show any clear singularity on the CVC.
In all studied cases instead of the periodic solution we have found oscillating,
but non-periodic solution. The Fourier transform of the voltage contains broad features instead of
$\delta$-function like spectrum before the transition.
The solution in the vicinity of $j_{c2}(L)$ is characterized by long intervals of time where the solution is
almost periodic and topologically similar to the solution before the bifurcation point $j<j_{c2}(L)$. Then the
instability develops and solution moves far away from the periodic cycle. After few periods the solution
is coming back again to a very similar oscillating trajectory and performs regular oscillations for many periods.
To demonstrate this behaviour we calculate Poincar\'e map for $j_{c2}<j<j_{c5}$ and $L=2L_{0}$. The projection of
the phase trajectory to the plane ($\rho(-L/4),E(-L/4)$) never crosses the space near the center of the trajectory. Therefore
we chose Poincar\'e section as a plane which crosses the center of this trajectory $E(-L/4)=0.04$ for $j/j_{c}=1.0278$. In Fig.(4) we plot
the projection of the Poincar\'e map to $\rho(0)$ and $E(0)$ as a function of discrete time, when trajectory crosses
the Poincar\'e section.
The long horizontal lines represent an
almost periodic solution. The time, during
which solution stays near the periodic laminar orbit becomes larger and larger when $j\to j_{c2}(L)+0$,
as demonstrated in the inset of Fig.4 \cite{Landau,Anischenko}. According to Eq. (32,23) of Ref.\cite{Landau} the
time, which trajectory stays near the regular limit cycle is proportional to
$\tau\propto (j-j_{c2}(L))^{-1/2}$. Inset in Fig.4 demonstrates that $\tau$ indeed increases as
$(j-j_{c2}(L))^{-1/2}$ indicating that we deal with chaotic solution which is developing via
the intermittence \cite{Anischenko}.

Further increase of the current for the case of $L=2L_{0}$ leads to an additional step of voltage
($j_{c3}(L)/j_{c}=1.058$) in the CVC. In that case two remote PSCs located symmetrically with respect
to the center of the channel are accompanied by the third one, which appears in the middle of the wire (Fig.2 d).
This does not destroy chaos, although changes apparently its topological properties.
The CVC in this region of currents have some non-regular behaviour (Fig.3 b),  which might be associated
with appearance and disappearance of the so-called ``periodic windows'', areas of the parameter values,
where limit cycles with a relatively large period become stable \cite{Anischenko}.


The next step in voltage at $j_{c4}(L)/j_{c}=1.084$ is associated with the
creation of a new PSC, again without the change of the chaotic character of the solution.
And finally, the non-periodic orbit
disappears and a limit cycle becomes stable in the relatively large window of currents
($j_{c5}(L)/j_{c}=1.098 <j/j_{c}<1.114$).
This bifurcation produces a voltage step on the CVC, and
in contrast to the all previous cases, the voltage decreases at
the bifurcation point (Fig. 3 b).
The corresponding limit cycle has four PSCs, each of which has different positions in the channel.
Remarkably, such behaviour of PSC was previously observed in experiments \cite{Sivakov}.

The situation for the longer channel $L=4L_{0}$ is slightly different. First, we have found that
for $j>j_{c1}(L)$ there are few ranges of the current where other limit cycles coexist with the
main periodic solution, similarly to the earlier report\cite{mic}. Therefore, at the same value of
the current we observe different stable periodic
solutions with their own attracting phase spaces. The multistability is found only in the case of
long channel ($L=4L_{0}$). In the case of $L=L_{0}$ and $2L_{0}$ we were unable to find any manifestation of multistability.
The region where chaos is developed via the intermittence for the long channel $L=4L_{0}$
is much more narrow then in the case of $L=2L_{0}$. The chaotic oscillations becomes very quickly unstable and
new stable limit cycle with 3 PSCs appears at $j/j_{c}>j_{c3}(L)/j_{c}=1.0011$ in contrast with the former case,
where this solution has also an irregular chaotic behaviour. This bifurcation manifests itself by the step in
voltage on the CVC and is consistent with the observation made in Ref.\cite{Sivakov}.

The effect of multistability revealed in our calculations may be related to the effects of switching between two
metastable states, discussed in Ref.\cite{Zybtsev}. Indeed the existence of two different limit cycles with the different topology
of phase trajectories and different averaged voltage implies that relatively large thermodynamic fluctuations may switch
from one trajectory to another. If the switching is fast and the averaged time between switchings is large in comparison with the
period of the limit cycle, the measured voltage will be averaged over many periods. The switching requires overcoming the energy
barrier between two limit cycles therefore the time between two switchings is exponential
function of temperature. In the case of wide bridges the activation energy is macroscopic and depends on the width of the sample. Therefore in the case when
the time of measurements is larger than the period of the cycle and shorter than the average time between the switchings
the fluctuations of the voltage looks like a telegraph noise and are similar to that observed in Ref.\cite{Zybtsev}.

In conclusion, we have analyzed the current-voltage characteristics of the superconducting channel of
a different length. We have proven that the
steady state solution losses its stability as a result of
the saddle-node homoclinic bifurcation, which leads to a limit cycle of the diverging period. This unambiguously leads
to the singular $V\propto (j-j_{c})^{1/2}$ CVC. We have demonstrated that the second anomaly in the CVC is caused
by the period
doubling bifurcation. The main consequence of this bifurcation is the appearance of the new frequency
in the spectrum of electromagnetic radiation which is equal exactly the half of frequency before bifurcation.
The appearance of the third PSC is not universal. In short channels this bifurcation appears when the solution has
chaotic character. Increasing the channel's length $L$ stabilizes a periodic solution with 3 PSC. In the
intermediate range of currents the periodic solution becomes unstable leading to the chaos, which is developed via
the intermittence. The effect of multistability in the presence of thermodynamic fluctuations may be responsible for the switching between
two periodic solutions and random jumps of voltage between two averaged values.

\begin{figure}
\includegraphics[width = 55mm, angle=-90]{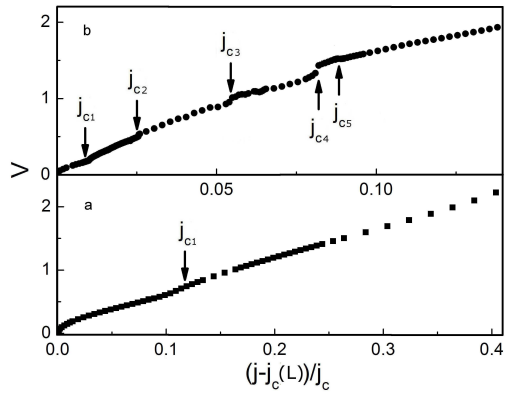}
\caption{The CVC of the wire a) $L=L_{0}$, b) $L=2L_{0}$. Arrows indicate different bifurcation points.}
\end{figure}

\begin{figure}
\includegraphics[width = 55mm, angle=-90]{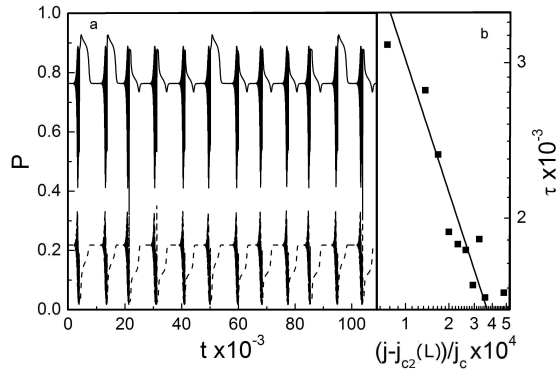}
\caption{a) The projection of the Poincar\'e map on $\rho(0)$ (solid line) and $E(0)$ (dashed line) as a function of time.
b) The averaged period of the laminar phase as a function of current (squares). The solid line
shows analytical prediction (see the text).}
\end{figure}

\end{document}